\documentclass[]{emulateapj}

\newcommand{\amm}{NH$_{3}$}
\newcommand{\mh}{H$_{2}$}

\newcommand{\cden}{cm$^{-2}$}
\newcommand{\vden}{cm$^{-3}$}

\newcommand{\msun}{M$_{\sun}$}
\newcommand{\kms}{km~s$^{-1}$}
\newcommand{\nw}{N\,159\,W}
\newcommand{\sfr}{M$_{\sun}$\,yr$^{-1}$}

\shorttitle{Ammonia in the LMC}
\shortauthors{Ott et al.}

\begin{document}

\title{First Detection of Ammonia in the Large Magellanic Cloud: The Kinetic Temperature of Dense Molecular Cores in \nw}

\author{J\"urgen Ott\footnote{Jansky Fellow of the National Radio Astronomy Observatory}}
\affil{National Radio Astronomy Observatory, P.O. Box O, Socorro, NM 87801, USA}
\affil{California Institute of Technology, 1200 E. California Blvd., Caltech Astronomy 249-17, Pasadena, CA 91125, USA}
\email{jott@nrao.edu}

\author{Christian Henkel}
\affil{Max-Planck-Institut f{\"u}r Radioastronomie, Auf dem H{\"u}gel 69, 53121 Bonn, Germany}
\email{chenkel@mpifr-bonn.mpg.de}

\author{Lister Staveley-Smith}
\affil{International Centre for Radio Astronomy Research, M468, University of Western Australia, 35 Stirling Highway, Crawley, WA 6009, Australia}
\email{Lister.Staveley-Smith@uwa.edu.au}

\and

\author{Axel Wei{\ss}}
\affil{Max-Planck-Institut f{\"u}r Radioastronomie, Auf dem H{\"u}gel 69, 53121 Bonn, Germany}
\email{aweiss@mpifr-bonn.mpg.de}

\begin{abstract}
  The first detection of ammonia (\amm) is reported from the
  Magellanic Clouds. Using the Australia Telescope Compact Array, we
  present a targeted search for the ($J$,$K$) = (1,1) and (2,2)
  inversion lines towards seven prominent star-forming regions in the
  Large Magellanic Cloud (LMC). Both lines are detected in the massive
  star-forming region \nw, which is located in the peculiar molecular
  ridge south of 30\,Doradus, a site of extreme star formation
  strongly influenced by an interaction with the Milky Way halo. Using
  the ammonia lines, we derive a kinetic temperature of $\sim16$\,K,
  which is 2-3 times below the previously derived dust
  temperature. The ammonia column density, averaged over
  $\sim17\arcsec$ is $\sim6\times10^{12}$\,\cden\ ($<1.5\times
  10^{13}$\,\cden\ over $9\arcsec$ in the other six sources) and we
  derive an ammonia abundance of $\sim4\times10^{-10}$ with respect to
  molecular hydrogen. This fractional abundance is 1.5-5 orders of
  magnitude below those observed in Galactic star-forming regions. The
  nitrogen abundance in the LMC ($\sim10$\% solar) and the high UV
  flux, which can photo-dissociate the particularly fragile \amm\
  molecule, must both contribute to the low fractional \amm\
  abundance, and we likely only see the molecule in an ensemble of the
  densest, best shielded cores of the LMC.
\end{abstract}

\keywords{galaxies: individual(Large Magellanic Cloud) --- ISM: molecules --- (galaxies:) Magellanic Clouds --- stars: formation ---  galaxies: ISM --- radio lines: ISM} 
\objectname{Large Magellanic Cloud}
\facility{Australia Telescope Compact Array}

\section{Introduction}
Our position within the disk of the Milky Way makes it very difficult
to understand galaxy--wide effects that may influence the formation of
molecular clouds, their stability against gravitational collapse, and
the subsequent formation of stars. Ideally, one would like to study a
more face-on, nearby galaxy and fortunately such objects exist, the
Magellanic Clouds. The Large and Small Magellanic Clouds (LMC, SMC)
are at distances of $\sim50$\,kpc and $\sim60$\,kpc, respectively
\citep[][this LMC distance is used throughout the
paper]{mad91,gie98,kel06}. The proximity is ideal for observations to
provide the high detail needed to study the formation and evolution of
giant molecular clouds and even to ``zoom'' into individual prestellar
cores (at the distance of the LMC 1\arcsec\ corresponds to
0.24\,pc). The LMC exhibits a favorable inclination of $\sim35\degr$
\citep{vdm01}. This value is ideal for both, a detailed census of all
molecular clouds and their distribution as well as to determine their
kinematics from radial velocities. The LMC may therefore be the very
best object to study the influence that galaxy-wide processes exert on
the star formation (SF) properties of an entire galaxy.

The ``fuel'' for SF is molecular gas. Good knowledge of its physical
and chemical condition, such as temperature and density, is
indispensable if we are to understand under which conditions, and
through which processes, and phases the molecular clouds collapse to
protostars. Molecular lines are a rich source for such
information. The sites of actual SF, the prestellar cores, are
typically very dense and cold and they may be surrounded by a warmer,
more tenuous molecular envelope. However, rotational lines from linear
molecules, such as the very abundant CO, have only limited power to
distinguish between the two phases. Radiative transfer models, such as
``Large Velocity Gradient'' (LVG) codes
\citep[e.g.][]{sco74,kwa75,vdt07}, show that cold and dense gas on one
hand, and hot and tenuous gas on the other hand, produce similar
rotational line ratios. This degeneracy can, in principle, be broken
by observing a large number of rotational lines, including a variety
of isotopomeres or even different molecular species
\citep[e.g.,][]{wei01a,wei07}. Such an approach requires extremely
well-calibrated spectra at millimeter and sub-millimeter wavelengths,
which are difficult to obtain. Different beam sizes lead to additional
complexity and resulting parameters also depend on the detailed
properties of the applied radiative transfer model.

A much more direct and therefore more reliable path is to observe
lines from molecules with non--linear structures. One of the most
abundant non--linear molecule in the Universe is ammonia (\amm). \amm\
is a tetrahedral symmetric--top with the exceptional ability for the
nitrogen atom to tunnel through the plane defined by the three
hydrogen atoms. This peculiar feature produces a series of energy
doublets, and the transitions between the levels of such pairs are
known as inversion transitions. As a symmetric-top, the level scheme
of \amm\ is not, as in the case of CO, characterized by a single
($K$=0) rotational ladder. Instead, there is a multitude of
$K$-ladders, whose relative populations are determined by
approximately Boltzmann-distributed collisional processes, providing
direct information on kinetic temperatures.

Despite previous searches for ammonia in the Magellanic Clouds
\citep[e.g.][]{ost97}, the molecule has not been detected in the
past. In this paper we provide deeper data and present the first
detection of ammonia in the Magellanic system.

\section{Observations and Data Reduction}
\label{sec:obs}
The observations, all performed with the Australia Telescope Compact
Array (ATCA\footnote{The Australia Telescope Compact Array is part of
  the Australia Telescope which is funded by the Commonwealth of
  Australia for operation as a National Facility managed by CSIRO.}),
were set up to accommodate two different ammonia search strategies,
both with the same primary beam (field of view) of 2\farcm4. One
strategy was to perform a survey (project code C1372) toward seven
different star-forming regions. The source names and positions are
listed in Table\,\ref{tab:pos} and were selected from samples of
particularly prominent, line-rich regions studied in
\citet{chi97} and \citet{hei99}. The second part of our strategy was
to obtain a deep integration (project code CX058) toward one of the
targets, the brightest molecular clump in the star-forming region
N\,159, N\,159-West (hereafter ``\nw''), which is known to be one of
the two most intense molecular hot spots of the Magellanic Clouds.

The multi-source survey was performed on 2005 March 19 and 20 when the
ATCA was in the H\,214 antenna configuration. The calibrators
PKS\,1934-638, PKS\,1921-293, and PKS\,0454-810 were used for flux,
bandpass, and gain/phase calibration, respectively. The integration
time on each position was about 180\,min split up into 9 different LST
intervals, with gain/phase calibrator scans in between, for better
{\sl uv}--coverage. The receivers were tuned to observe the \amm(1,1)
line at high resolution, centered on 23.673\,GHz, with a bandwidth of
8\,MHz and 512 channels. This sky frequency was chosen to accommodate
a velocity range of $\sim85$\,\kms\ for the (1,1) line, centered on an
LSR velocity of 255\,\kms\ (rest frequency for the [1,1] line:
23694.4955\,MHz, for the [2,2] line: 23722.6333\,MHz\footnote{taken
  from the JPL Molecular Spectroscopy Database: {\it
    http://spec.jpl.nasa.gov}}). The almost circular synthesized beam
varies slightly from position to position but the linear resolution
hovers around $\sim9\arcsec\pm1\arcsec$ FWHM (for the sensitivity, see
Sect.\,\ref{sec:results}).

The deeper, targeted ammonia observations toward \nw\ were executed
during three observing periods, 2006 April 27 ($\sim5$ hours
integration time, H\,214 array configuration), 2006 September 15
($\sim4$ hours, H\,75), and 2006 September 16 ($\sim2.5$ hours,
H\,75). We used the same flux and complex gain calibrators as before,
but the bandpass was calibrated using observations of
PKS\,0537-441. To accommodate the \amm\ (1,1) and (2,2) transitions
simultaneously, we tuned the two available frequencies of the ATCA to
23.675\,GHz and 23.703\,GHz, respectively. The setup was not symmetric
but has 8\,MHz bandwidth and 512 channels for \amm\ (1,1)
(corresponding to 85\,\kms\ velocity coverage and a resolution of
$\sim0.2$\,\kms), and 16\,MHz bandwidth with 256 channels for \amm\
(2,2) ($\sim170$\,\kms\ with $\sim0.8$\,\kms\ resolution). To have
consistent data cubes, we binned both uv data sets to the same channel
spacing of $\sim0.8$\,\kms. The rms noise per channel map is $\sim
2$\,mJy within a $18\farcs6\times15\farcs7$ (PA$=74\degr$) synthesized
beam. This corresponds to a brightness temperature of $\sim
15$\,mK. For a better representation of the lines, we boxcar-smoothed
the binned spectra with a width of five channels in velocity space. The
smoothed data, which widens the true velocity widths by a factor
of $\sim\sqrt{2}$ (see also Sect.\,\ref{sec:mass}), are used for all
further analysis.

\section{Results}
\label{sec:results}

\begin{figure} 
\epsscale{1} 
\plotone{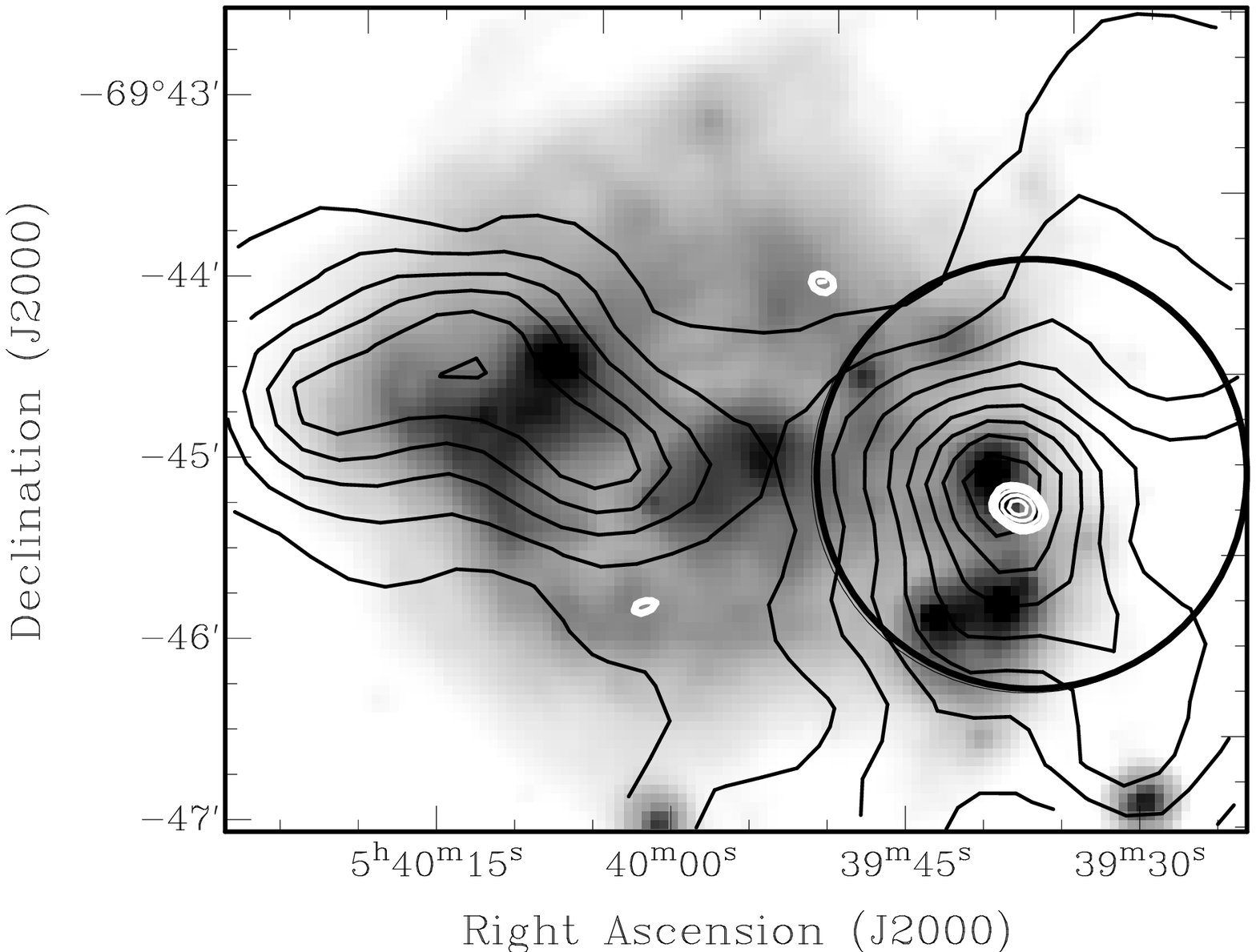} 
\caption{Integrated ammonia (1,1) contours ({\it white}) on a Spitzer
  MIPS 24\,$\mu$m image ({\it greyscale}, logarithmic scaling). The
  contours start at 0.15\,K\,\kms\ ($\sim3\sigma$, thick white
  contour) and are spaced by 0.05\,K\,\kms\ (thinner white
  contours). The {\it black circle} indicates the 2\farcm4 primary
  beam of the ATCA \amm\ observation. For comparison, we also show the
  $^{12}$CO($1\rightarrow0$) distribution in {\it black contours} as
  observed with the Mopra telescope at $\sim45\arcsec$ resolution
  \citep[data taken from][]{ott08}. The strong IR source north of the
  ammonia emission is known as \nw\ AN. \label{fig:map}}
\end{figure}

\begin{figure} 
\epsscale{1} 
\plotone{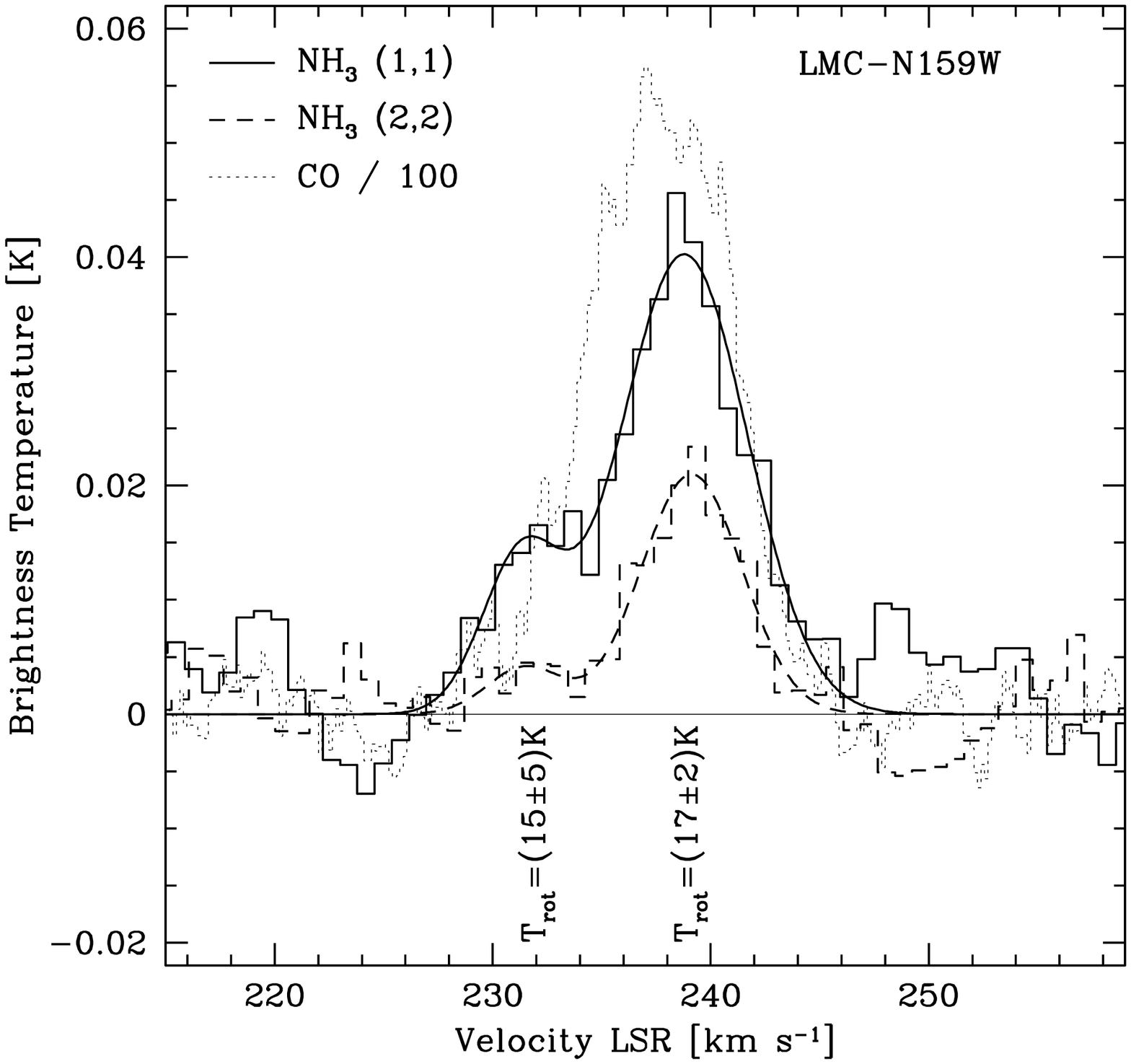}  
\caption{Smoothed \amm\ (1,1) ({\it solid lines}) and (2,2) ({\it
    dashed}) spectra toward \nw\ ($\alpha[J2000]=05^{h}39^{m}36\fs1,
  \delta[J2000]=-69\degr45\arcmin38\arcsec$) and their respective
  two-component Gaussian fits. A CO spectrum of the position (obtained
  within a beam of $\sim45\arcsec$, and scaled down by a factor of
  100) is shown as a {\it dotted line}. \label{fig:spec}}
\end{figure}

The survey toward the different regions results in non-detections with
an rms of $\sim$4\,mJy in a 0.8\,\kms\ channel. With the rather small
synthesized beam of $\sim9$\arcsec, this translates to a brightness
temperature of $\sim120$\,mK. The deeper (in terms of brightness
temperature) integration towards \nw, however, clearly reveals the
\amm\ (1,1) line which breaks up into two different velocity
components. The \amm\ (2,2) line is detected with a lower statistical
significance, in particular the low-velocity component. An overlay of
the integrated intensity of the \amm (1,1) line on a Spitzer MIPS
24\,$\mu$m map is shown in Fig.\,\ref{fig:map}. The ammonia emission,
centered on $\alpha_{\rm 2000}=05^{h}39^{m}36\fs1, \delta_{\rm
  2000}=-69\degr45\arcmin38\arcsec$, coincides with a peak seen in the
$J=1\rightarrow0$ transition of carbon monoxide (CO data, at a
resolution of $\sim45\arcsec$, taken from \citealt{ott08}, see also
\citealt{pin09}), and is located very close to the bright IR source
\nw\ AN \citep[e.g., see][]{jon05}. The ammonia emitting region has a
very similar size and axis ratio as the beam of the observations,
which indicates that it is unresolved. We also do not expect to lose
extended flux by the spatial filtering of the interferometer as we
observed in the most compact H75 antenna configuration of the ATCA. At
the observed frequencies, this antenna configuration is sensitive to
scales up to 80\arcsec, or 20\,pc at the distance of the LMC (about
4-5 times the size of the synthesized beam). However, a visual
inspection indicates that the position angle may deviate marginally
from that of the beam (see Sect.\,\ref{sec:obs}). In the following, we
assume that the emission is not resolved and leave a better
characterization of the source structure to deeper observations at
better spatial resolution. The \amm\ (1,1) and (2,2) spectra are
displayed in Fig.\,\ref{fig:spec}. Gaussian fits with a low- and a
high-velocity component provide good descriptions for the individual
line profiles. The fit parameters are listed in
Table\,\ref{tab:params}. The high, $\sim239$\,\kms\ velocity
component is about a factor of $\sim3$ stronger than the low, $\sim
231$\,\kms\ velocity component and it is slightly broader.

We do not detect the typical ammonia satellite lines at offset
velocities of $\sim\pm 8$\,\kms\ and $\sim\pm 20$\,\kms. The strength
of the satellite lines with respect to the main line depends on the
optical depth of the line. \cite{ho83} find that the peaks of the
inner satellite lines with respect to the peak of the main line are
described by $T_{\rm mb, peak} (\rm main\; lines)/T_{\rm mb, peak}
(\rm inner\; satellite\;
lines)=(1-\exp[-\tau])/(1-exp[-0.28\,\tau])$. For the extreme case of
optical thin lines ($\tau<<1$) this ratio approaches $\sim3.6$; in the
optically thick case the ratio would be unity. At $\sim 248$\,\kms\ we
detect a faint feature that could correspond to an inner satellite
line in the optically thin limit. Its significance over the smoothed
data, however is only $\sim2.5\sigma$ (note that the symmetric line
blends with the low-velocity component; a feature at $\sim220$\,\kms\
may indicate an outer satellite line). A clear, $4\sigma$ detection of
an inner satellite would correspond to an opacity of
$\tau\sim1.3$. This conservative upper limit allows thin and moderate
opacities for the ammonia emitting gas and we are able to exclude the
optically thick case. While we cannot prove it, we assume that the
optical depth of the other, weaker \amm\ velocity component also is
only moderately optically deep at most.

\placefigure{fig:map}

\placefigure{fig:spec}

\placetable{tab:pos}

\placetable{tab:params}

\section{Discussion}
\label{sec:discuss}

\subsection{The Kinetic Temperature of  \nw}
\label{sec:T}

Assuming the optically thin case (see above), the column densities of
the upper ammonia levels of the metastable ($J$,$J$) inversion
doublets, $N_{\rm u}$($J$,$J$) (in units of \cden), can be derived
using

\begin{equation}
N_{u}(J,J)=\frac{7.77\times10^{13}}{\nu}\frac{J(J+1)}{J^{2}}\,\,\,\int T_{\rm mb}\,\, dv
\label{eq:n}
\end{equation}
\citep[e.g.][with $T_{\rm mb}$ denoting the main beam brightness
temperature of the $(J,J)$ inversion line in Kelvin, $v$ being the
velocity in \kms, and $\nu$ representing the frequency in
GHz]{hen00}. Our values for \nw\ are listed in
Table\,\ref{tab:params}. For a single ortho- or para- ammonia species
(the observed lines in \nw\ all belong to para-ammonia) the Boltzmann
distribution for a rotational temperature $T_{\rm rot,JJ'}$ (in K) is described by

\begin{equation}
\frac{N_{u}(J',J')}{N_{u}(J,J)}=\frac{2J'+1}{2J+1}\,\exp\left(\frac{-\Delta E}{T_{\rm rot,JJ'}}\right).
\label{eq:t}
\end{equation}
Here, the energy difference between the $(J,J)$ and $(J',J')$ level is
given by $\Delta E$ [41.2\,K between \amm\ (1,1) and (2,2)]. We derive
a rotational temperature $T_{\rm rot,12}$ of $(17\pm2)$\,K for the
high-velocity \amm\ component and a very similar $(15\pm5)$\,K for the
low-velocity component (uncertainties are based on the $\pm1\sigma$
values for $N_{\rm u}$ derived in Eq.\,\ref{eq:n}, not quadratically
propagated but using the extremes). Rotational temperatures are always
lower limits to the kinetic gas temperatures. However, for
temperatures $\lesssim20$\,K the differences between $T_{\rm kin}$
and $T_{\rm rot,12}$ are less than 10\% and thefore negligible
\citep[e.g.][]{wal83,dan88}. We can safely assume $T_{\rm kin} =
T_{\rm rot,12}$ for both components of \nw.

Thermal linewidths of gas at such cold kinetic temperatures are
$\sim$0.2\,km\,s$^{-1}$. The measured \amm\ linewidths are at least an
order of magnitude larger than this figure and deconvolving the lines
by the instrumental velocity resolution and the applied boxcar
smoothing is not sufficient to reach the predicted thermal widths. The
gas kinematics must thus be dominated by other processes such as
turbulence or by systematic motions of spatially unresolved cloud
components. This is supported by the line widths of other molecules
which are similar to those of \amm\ \citep[e.g.][]{hei99}.

The kinetic temperature of the gas traced by \amm\ is colder than the
30-40\,K dust temperature that is measured in \nw\
\citep[][]{hei99,bol00}. In their LVG analysis of high excitation
molecular gas, \citet{bol05} establish a two-component model for \nw,
which comprises a gas phase with $\sim20$\,K kinetic temperature and a
volume density of $\sim10^{5}$\,\vden, as well as a second phase with
$\sim100$\,K and $\sim10^{2}$\,\cden. Likely, the detected ammonia
traces the low temperature component, albeit at larger volume
densities (for volume density estimates based on \amm, see
Sect.\,\ref{sec:mass}). This could be in the form of cold cores
surrounded by a warmer envelope. As mentioned above, dust temperatures
in the region have been estimated to be $\sim2-3$ times that of \amm\
and it is likely that the dust is not only present in the cold cores
but also occupies the interface to the warm envelope, covering a
rather wide range of densities.

The detailed dust and gas morphologies have implications for the
shielding against UV radiation. The \amm\ molecule is quite
susceptible to photo-dissociation \citep{sut83}. \nw\ itself is a
rather strong source of ionizing photons \citep[e.g.][also see
Sect.\,\ref{sec:sf}]{jon05} and a smaller contribution comes from the
most luminous star-forming region in the Local Group, 30\,Doradus,
which is about $\sim600$\,pc away from \nw\ \citep[see,
e.g.][]{pin09}. The flux of energetic UV photons may penetrate into
the envelope and destroy ammonia. It requires dense cores to provide
enough shielding to keep ammonia intact and it is therefore likely
that the molecule resides within such dense, cold, compact cores. In
addition, unlike CO, nitrogen-bearing species such as \amm\ and
N$_{2}$H$^{+}$ are less prone to depletion on dust grains at densities
typical for cold dense cores such as protostellar cores
\citep[$>10^{5}$\,\vden; e.g.][for a density estimate of \nw\ see
Sect.\,\ref{sec:mass}]{taf04,flo05} and \amm\ is expected to be
abundant in such objects.

\subsection{Densities, Masses, and Abundances}

\label{sec:mass}

For a given single rotational temperature, total \amm\ column
densities, $N({\rm NH_{3}}, \rm total)$, can be derived by
extrapolating the level distribution over all energies. \citet{ung86}
provide this extrapolation in their Eq.\,A15, which only requires
$T_{\rm rot, 12}$ and $N_{\rm u}(1,1)$ to be determined. For \nw\ we
derive a total, 17\arcsec\ beam-averaged, ammonia column density of
$(5.8\pm0.6)\times10^{12}$\,\cden. Toward the same position, the
integrated CO($1\rightarrow0$) intensity is $\sim45$\,K\,\kms\ over a
45\arcsec\ beam \citep[Mopra data taken from][see also
Fig.\,\ref{fig:map}]{ott08}. Assuming a CO to \mh\ conversion factor
of $X_{\rm CO}\sim4\times 10^{20}$\,\cden\,(K\,\kms)$^{-1}$ for the
LMC \citep{pin09}, the \mh\ column density can be estimated to be
$\sim1.8\times10^{22}$\,\cden. With this value, the fractional ammonia
abundance in the \nw\ complex becomes $\sim 3.3\times10^{-10}$. This
number is based on a 45\arcsec\ average of the \mh\ column
density. The \mh\ distribution within that area, however, is most
likely not smooth but exhibits some variation. To excite \amm, a
critical density needs to be exceeded. We do not observe \amm\ outside
the smaller, 17\arcsec\ ATCA beam, which indicates that the \mh\
density is likely higher at the position of the \amm\ emission than
further away from it. Consequently, the ammonia abundance of $\sim
3.3\times10^{-10}$ may be considered an {\it upper} limit and any \mh\
column exceeding the average over 45\arcsec\ will dilute the ammonia
abundance. Furthermore, larger $X_{\rm CO}$ factors, such as reported
by \citet{isr03} or \citet{fuk08}, would also drive the fractional
ammonia abundance to smaller values.

The $X_{\rm CO}$ factor is just one way to estimate the total \mh\
content of the molecular cloud. A different approach is to derive
virial masses $M_{\rm vir}$, assuming that the system is virialized
and the masses of the clouds are dominated by \mh. For a cloud with a
Gaussian density profile, the virial mass can be estimated using

\begin{equation}
M_{\rm vir}=444\,r\,\Delta v_{\rm 1/2}^{2}
\label{eq:virial}
\end{equation}
\citep[][]{pro08} with the radius $r$ in pc, the full width at half
maximum (FWHM) velocity $\Delta v_{\rm 1/2}$ in \kms, and the virial
mass $M_{\rm vir}$ in M$_{\sun}$ \citep[note that the numerical factor
changes with the shape of the radial density profile; e.g. it is $\sim
2$ times lower for $\rho=\rm constant$, and $\sim3.5$ times lower for
a $\rho\propto r^{-2}$ profile;][]{mla88}. The \amm\ emission position
angle may show some deviation from a pure point source but, given the
faintness of the source, we use half of the FWHM ATCA beam size as an
upper limit to the radius of the cloud, 8\arcsec\ which corresponds to
$\sim2$\,pc. Together with $\Delta v_{\rm 1/2}\sim5$\,\kms, the
virial mass amounts to an upper limit of $<2.2\times10^{4}$\,\msun\
(for comparison, the virial mass based on the Mopra CO data over the
much larger Mopra beam would be $\sim10^{5}$\,\msun). This limit also
accounts for the boxcar smoothing of the data; a more narrow line
would lower the upper limit and our estimate is therefore a more
conservative one. In addition, this estimate also accounts for
material other than \mh\ that is contributing to the virial mass,
justifying the upper limit nature of the \mh\ mass based on the virial
method.

Using a spherical geometry of the \amm\ emission and assuming that all
the virial mass is due to \mh, the mean volume density of the
molecular gas then can be estimated to $\lesssim
1.3\times10^{4}$\,\vden. This value is fairly typical for dense gas in
Galactic star-forming regions, spread over the physical size of the
beam, $\sim4$\,pc \citep[e.g.][]{fos09}. Finally, if we take the
column density of ammonia of $(5.8\pm0.6)\times10^{12}$\,\cden\ (see
above), we derive a beam averaged ammonia mass of $\sim
1\times10^{-5}$\,\msun, or $\sim3.3$\,M$_{\earth}$. With \mh\ masses
and densities being upper limits, the fractional abundance of ammonia
(molecule number density ratio) then follows as a {\it lower limit} of
$\gtrsim4.5\times 10^{-10}$. This value is in agreement with the upper limit
derived via $X_{\rm CO}$ above.

We cannot infer any robust variation in fractional ammonia abundance
between the two velocity components. The CO(1$\rightarrow$0) emission
is about two orders of magnitude brighter (cf. Fig.\,\ref{fig:spec})
at the velocity of the ammonia high-velocity component. The ratio
appears to be somewhat smaller for the low-velocity component, which
would indicate some variation in the abundance. But, given the
uncertainties in beam filling factors for the CO and \amm\
observations (see also Table\,\ref{tab:params}), better data is needed
to reliably establish such a variation.

Using the above values for temperature and upper volume density limit
for \nw, we estimate the pressure of the clumps in the \nw\ region to
be $P/k\gtrsim2\times10^{5}$\,K\,\vden. This value is likely to
increase substantially when the average value for \mh\ densities is
resolved into individual, smaller clumps with relatively little \mh\
in between. For comparison, \citet{won09} calculate the midplane
pressure for the positions where molecular gas is observed in the
LMC. Their highest value is close to our lower limit, which may
indicate that the dense \amm\ clumps are not in pressure equilibrium
with their surroundings, a property that is typical for
self-gravitating clouds. We like to note though that the volume
density of our calculation is actually based on the virial theorem,
i.e. a gravitational equilibrium of the molecular gas that is
independent of its environment.

We can now revisit the upper limits of the ammonia survey toward the
other six positions. If we assume the same linewidth and kinetic
temperature that we find for the two components of \nw, the $3\sigma$ upper
limits to the non--detections translate to ammonia columns $N_{\rm u}
(1,1)$ and $N ({\rm NH_{3}}, \rm total)$ of $<5\times10^{12}$\,\cden,
and $<1.5\times10^{13}$\,\cden\ for a $9\arcsec$-sized region,
respectively.

\subsection{Star Formation Properties of \nw}

\label{sec:sf}

If we take the upper limit to the virial mass of the ammonia emitting
region of $\sim2\times10^{4}$\,\msun\ for the mass of the molecular
clump (see Sect.\,\ref{sec:mass} above), the typical SF efficiency of
a few percent \citep[][]{eva09} would convert the gas to
stellar masses of $\sim10^{3}$\,\msun. This falls within the mass
range of open stellar clusters. To be independent of extinction
effects \citep[the very nearby ionizing source \nw\ AN is deeply
enshrouded by dust as shown by][]{jon05}, we use the 1.4\,GHz radio continuum map
presented in \citet{hug07} to estimate the current SF rate
(SFR) via the conversions provided by \citet{haa00}. For \nw\ AN, we
measure a flux of $\sim260$\,mJy which corresponds to a current SFR
of $\sim8\times10^{-4}$\,\sfr. The radio emission at that position,
however, cannot be entirely separated from other nearby sources and we
estimate an uncertainty to the flux and the SFR of up to 50\%. If 
SF continues at that level, the available molecular gas will be fully
converted into stars within the next $\sim10$\,Myr.

\subsection{Comparison of  \nw\ with Galactic and Extragalactic Star Forming Regions}

\label{sec:galactic}

The LMC has a lower mass and metallicity than the Galaxy. In addition,
the \nw\ region is located in the molecular ridge south of the
extremely active 30\,Doradus star-forming region, at a location where
the LMC's regular disk becomes disturbed by tidal and ram pressure
forces that act as the galaxy moves through the Milky Way halo
\citep[e.g. ][]{ott08}. Do these peculiarities change the properties
of the molecular clumps formed at this location? In order to address
this question we compare our results to the properties of dense
molecular cores in the Galaxy. A homogenized sample of molecular
cores, drawn from Galactic star-forming regions such as Orion,
Ophiuchus, Perseus, Taurus, Serpens, Cepheus, etc. is presented in
\citet{jij99}. The authors have collected ammonia (1,1) and (2,2)
observations for all of their sample and can therefore provide data
that may act as a reference for comparison with clouds in other
galaxies such as the LMC. More recently, \citet{ros08} \citep[see
also][]{stu92,fos09} published a comprehensive ammonia survey of 193
dense molecular cores toward the Perseus molecular cloud, a cloud that
may also be representative for typical star-forming regions in the
Galaxy. The studies agree that Galactic dense molecular cores
exhibit typical kinetic gas temperatures of $\sim10$\,K, with sizes
of $\sim0.1$\,pc, velocity widths of $\sim0.1-0.7$\,\kms, ammonia
column densities of $\sim10^{14.5}$\,\cden, and virial masses of
$\sim5$\,\msun. In most aspects these values are clearly different
from what we measure toward the LMC cloud \nw. The difference can be
explained by observing an ensemble of cores within the ATCA beam. Such
a site could potentially be the birth of an entire stellar cluster
(see also Sects.\,\ref{sec:mass} and \ref{sec:sf}). Along these lines,
the \nw\ region may consist of individual, dense molecular cores that
are at temperatures very similar to those found in the Galaxy 
  along with some inter-clump material. Together, the line profiles
of the clumps and the more diffuse gas blend to form the
measured ammonia lines with a width of $\sim5$\,\kms\ (note that
\citealt{ros08} also largely attribute the upper end of their measured
linewidths to blending of a number of cores along the line of
sight). We would like to note that the \nw\ SF timescales as derived
in Sect.\,\ref{sec:sf} also match those that are typical in Galactic
SF regions \citep{eva09}.


The differences in ammonia column densities and fractional abundances
between Galactic cores and the emission in \nw, however, cannot be
reconciled by beam filling arguments. The \amm\ column densities in
\nw\ are about 50 times lower ($\log(N[NH_{3}])\sim12.8$ for \nw,
$\sim14.5$ for Galactic SF regions). If beam filling would be
responsible for the difference, the ATCA beam would be diluted by
about the same factor. This implies that of the 20\,pc$^{2}$ ATCA beam
area only 0.4\,pc$^{2}$ is covered by clouds. If these clouds are
similar to Galactic cores (with typical sizes of 0.1\,pc and masses of
$\sim5$\,\msun, see above), the coverage corresponds to $\sim50$
molecular cores adding up to a total mass of $\sim250$\,\msun. In
Sect.\,\ref{sec:mass} we derive an upper limit of $\sim2\times
10^{4}$\,\msun\ to the virial mass in \nw, which is about two orders of
magnitude larger. This indicates that the basic assumption for the
calculation, the equality of Galactic and \nw\ ammonia column
density, is wrong.

The virial mass is an upper limit due to unknown mass components other
than \mh\ within the beam and due to the fact that the ATCA beam was
taken as the virial radius of the unresolved emission. Can the
parameters play together to create the $250$\,\msun\ mass as predicted
from Galactic clumps? We measure a velocity width of 5\,\kms\ within
\nw\ and, according to Eq.\,\ref{eq:virial}, this determines the size
of the gaseous emission for the $250$\,\msun\ to be $0.05$\,pc if the
clouds are in virial equilibrium. Such a small value is clearly
unrealistic and we conclude that it is not possible to reconcile the
ammonia column densities with those of Galactic SF regions based on
beam filling arguments alone. It is likely that intra-clump gas covers
a larger region of the beam and/or a larger number of cores is present
in \nw. But this implies that the ammonia column densities in \nw\ are
intrinsically lower than those in Galactic cores.

Similarly, the low fractional abundances of ammonia in \nw\ cannot be
fully explained by clumping of the gas within the beam (note that the
the abundances are independent of the beam size when applying the
virial mass as in Sect.\,\ref{sec:mass}). \citet{ben83} derive for
Galactic Dark Clouds on average fractional ammonia abundances of $\sim
10^{-7}$ and \citet{mau87} obtain $\sim10^{-5}$ for a hot core. The
resolved Galactic clumps as listed in the more recent \citet{ros08},
\citet{fos09}, and \citet{jij99} studies have \amm\ abundances of
order $\sim10^{-8}$. In contrast, we estimate a fractional ammonia
abundance in \nw\ of $\sim4\times10^{-10}$ (see
Sect.\,\ref{sec:mass}). \nw\ thus exhibits at least $\sim1.5$ orders of
magnitude lower fractional ammonia abundance than seen in individual
cores embedded in Galactic clouds. But \nw\ is not an exception: the
upper limits towards the other LMC star-forming regions in our survey
exhibit a limit of at least one order of magnitude below the Galactic
ammonia abundances. Such a difference cannot be an effect of the
overall lower metallicity of the LMC alone, which is $\sim45$\% solar
\citep[based mostly on iron, carbon, oxygen and
silicon]{luc98,rol02}. A combination of the $\sim5$ times lower
abundance of nitrogen in the LMC, $\sim 10$\% solar \citep{hun07}, and
the high UV flux at the \nw\ location \citep[see Sec.\,\ref{sec:T} and
also][]{deh92,bol00,pin09} that can potentially photo-dissociate the
fragile \amm\ \citep{sut83} must therefore contribute to its low
fractional abundance.

We cannot exclude that the fraction of clump to interclump
\amm\ emission might be very different in \nw\ as compared to Galactic
SF regions. In such a scenario, hardly any of the molecular gas
parameters can be reconciled between the two galaxies. Beam dilution
effects can explain most of the differences, except for the ammonia
abundance and column density, and is therefore a more natural
interpretation. To fully resolve this issue, sensitive observations at
better spatial resolution are required.

Among extragalactic objects, only M\,82 is known to exhibit a
similarly low fractional ammonia abundance like \nw. \citet{wei01b}
derive a value of $\sim5\times10^{-10}$ for this prominent starburst
galaxy. The gas in M\,82, at $T_{\rm kin}\approx 60$\,K, is warmer
than in \nw\ and the metallicity of M\,82 may also exceed that of the
LMC, in particular close to the central starburst region \citep[see,
e.g.][]{mcl93,mar97}. The presence of nearby, strongly ionizing
sources in both M\,82 and \nw\ suggests that photo-dissociation plays
an important role in regulating the ammonia abundances in both
galaxies.

\section{Summary}
\label{sec:summary}

In this paper, we present a search for ammonia toward seven prominent star
forming regions as well as deep observations towards the star-forming
\nw\ region close to 30\,Doradus. Observing the ($J$,$K$)=(1,1) and (2,2)
\amm\ inversion lines we find the following:

\begin{itemize}

\item Two ammonia velocity components are observed toward \nw, both
  with linewidths of $\sim5$\,\kms. The components are $\sim 8$\,\kms\
  apart. At least for the brighter line, we can exclude large optical
  depths. The linewidths and derived virial masses (of order $\sim
  10^{4}$\,\msun) indicate that we may see the molecular gas that
  forms an ensemble of stars similar to open clusters in the Galaxy.

\item We derive the kinetic temperature of the gas to be $\sim16$\,K
  for both components, $\sim$2-3 times colder than the dust in the region.

\item The ammonia column densities in all of the observed LMC
    star-forming regions are lower than typical values in our Galaxy,
    even after correcting for beam dilution effects.

  \item In the absence of a large, diffuse and hot component in \nw,
    the fractional ammonia abundance relative to \mh\ is with
    $\sim4\times10^{-10}$ within a $17\arcsec$ ($\sim4$\,pc) beam,
    about 1.5-5 orders of magnitude lower than in Galactic
    star-forming clouds and prestellar cores. The difference in
    abundances (based on this cloud) is therefore much larger than the
    metallicity difference of the LMC with respect to the Galaxy (LMC
    metallicity: $\sim45$\% solar). \amm\ photo-dissociation from the
    high UV flux in the LMC as well as its lower nitrogen abundance
    ($\sim10$\% solar) potentially explain the \amm\ deficiency.

\item If the gas is converted into stars with the current, radio
  continuum-based star formation rate, SF will cease within the
  typical SF timescale of a few Myr.

\item The multi-position survey did not reveal any detection down to a
  limit of $\sim120$\,mK in a 0.8\,\kms\ velocity bin. If there are
  molecular clumps with properties similar to those in \nw, we derive
  $3\,\sigma$ upper limits to the total column densities at these
  positions to be $<1.5\times10^{13}$\,\cden\ within a 9\arcsec\ (2\,pc)
  beam.

\end{itemize}

The Magellanic Clouds are unique as they allow us to perform analyses
similar to that possible for Galactic star forming regions, but
without the confusion that occur in edge-on systems. They are also
ideal to trace the effects that low metallicity has on the state and
chemistry of the dense molecular gas, the material that eventually
forms the stars. The kinetic temperature of the gas is a very
important parameter to understand the dense gas. It will be the focus
of future studies to extend the use of ammonia as a thermometer for a
larger number of gas clumps in order to obtain a more representative
sample.

\acknowledgements This work has been supported by the National Radio
Astronomy Observatory (NRAO) which is operated by Associated
Universities, Inc., under cooperative agreement with the National
Science Foundation. We like to thank the referee for his/her
suggestions to improve the paper.
 

\newpage

\begin{deluxetable}{lcc}
  \tablecaption{Target names and positions for the LMC-wide search for
    ammonia.\label{tab:pos}}
\tablehead{Name & RA (J2000) & DEC (J2000)}
\startdata
N\,83\,A & $04^{h}54^{m}01\fs68$ & $-69\degr11\arcmin28\farcs7$\\
N\,105\,A & $05^{h}09^{m}50\fs04$ & $-68\degr53\arcmin12\farcs7$\\ 
N\,113 & $05^{h}13^{m}18\fs20$ & $-69\degr22\arcmin24\farcs7$\\
N\,44\,BC & $05^{h}22^{m}01\fs26$ & $-67\degr57\arcmin36\farcs8$\\
30\,Dor\,10 & $05^{h}38^{m}50\fs93$ & $-69\degr04\arcmin17\farcs2$\\
N\,159\,W & $05^{h}39^{m}36\fs03$ & $-69\degr45\arcmin24\farcs7$\\
N\,159\,S & $05^{h}39^{m}59\fs29$ & $-69\degr50\arcmin23\farcs4$\\
\enddata
\end{deluxetable}

\begin{deluxetable}{ll|cc|cc}
\tablecaption{Parameters of the ammonia detection toward \nw. \label{tab:params}}

\tablewidth{0pt}
\tablehead{ & &\multicolumn{2}{c|}{Component 1} & \multicolumn{2}{c}{Component 2}\\parameter & unit &\amm\ (1,1)& \amm\ (2.2) & \amm\ (1,1) & \amm\ (2,2)}
\startdata
$T_{\rm mb, peak}$ & [mK] & $40.2\pm3.1$ & $21.0\pm1.$5 & $13.8\pm4.0$ & $4.1\pm1.8$\\
$v_{\rm peak}$ & [\kms] & $238.8\pm0.3$ & $239.1\pm0.2$ & $231.3\pm0.8$ & $231.5\pm0.9$ \\
$\Delta v_{\rm 1/2}$ & [\kms] & $6.8\pm0.8$ & $5.4\pm0.5$ & $4.3\pm1.7$ & $4.0\pm2.2$ \\
$\int T\,dv$ & [K\,\kms] & $0.29\pm0.04$ & $0.12\pm0.02$ & $0.06\pm0.03$ & $0.02\pm0.01$\\
$N_{\rm upper}$ & [$\times 10^{11}$\,\cden] & $19.2\pm2.8$ & $6.0\pm0.7$ & $4.1\pm2.0$ & $0.8\pm0.6$ \\
\hline
$T_{\rm rot,12}\approx T_{\rm kin}$ & [K] & \multicolumn{2}{c|}{$17\pm2$} & \multicolumn{2}{c}{$15\pm5$} \\
$N({\rm NH_{3}}, \rm total)$ & [$\times 10^{12}$\,\cden] &   \multicolumn{2}{c|}{$4.7\pm5$}  &   \multicolumn{2}{c}{$1.1\pm3$} 
\enddata
\end{deluxetable}

\end{document}